\documentclass[journal]{IEEEtran}
\usepackage{cite}
 \usepackage[pdftex]{graphicx}
\begin{document}
\title{ A Modular Data Acquisition System using the 10 GSa/s PSEC4 Waveform Recording Chip} 
%
%
%

\author{Mircea Bogdan,  Eric Oberla,
        Henry J. Frisch~\IEEEmembership{Member,~IEEE} and~Matthew
        Wetstein 
\thanks{M. Bogdan, E. Oberla, and H.J. Frisch are at the 
Enrico Fermi Institute, University of Chicago, IL 60637.}%
\thanks{M. Wetstein is at the Dept. of Physics, Iowa State
        University, Ames IA, 50011}
\thanks{Manuscript    received May 30, 2016}}

\maketitle

\begin{abstract}
We describe a modular multi-channel data acquisition system based on
the 5-15 Gigasample-per-second waveform-recording PSEC4 chip.  The
system architecture incorporates two levels of hardware with
FPGA-embedded system control and in-line data processing. The
front-end unit is a 30-channel circuit board that holds five PSEC4
ASICs, a clock jitter cleaner, and a control FPGA. The analog
bandwidth of the front-end signal path is 1.5 GHz.  Each channel has
an on-chip threshold-level discriminator that is monitored in the
FPGA, from which a flexible on-board trigger decision can be
formed. To instrument larger channel counts, a `back-end' 6U VME32
control card has been designed. Called the 'Central Card', it
incorporates an Altera Arria-V FPGA that manages up to 8 front-end
cards using one or two CAT5 network cables per board, which transmits
the clock and communicates data packets over a custom serial protocol.
Data can be read from the Central Card via USB, Ethernet, or dual SFP links,
in addition to the VME interface. The Central Card can be configured
as either Master or Slave, allowing one Master to receive data from up
to 8 Slaves, with each Slave managing 8 30-channel front-end cards,
allowing a single VME crate to control up to 1920 channels of the
PSEC4 chip.
\end{abstract}


%
\IEEEpeerreviewmaketitle

\section{Introduction}
The LAPPD Collaboration was formed in 2009 to develop detectors with
psec (10$^{-12}$ s)  resolution for particle physics
experiments~\cite{history_paper,Minot_NIM}. A necessary, but
non-existent, ingredient was the availability of suitable electronics with
resolution measured in psec, small footprint, parallelism measured in
thousands of channels, low-power, and low cost.

The LAPPD Collaboration consequently developed the PSEC4 ASIC, a
6-channel 5-15 GSa/sec waveform sampling chip with an analog bandwidth of
1.5 GHz, implemented in 130~nm IBM~8RF CMOS process~\cite{PSEC4_paper}.

In this paper we describe a modular data acquisition system (DAQ)
built around the PSEC4 chip.  The system architecture incorporates two
levels of hardware: the front-end unit is a 30-channel circuit board
that holds five PSEC4 ASICs. The back-end card incorporates an Altera
Arria-V FPGA that manages up to 8 front-end cards.  Data can be read
off the board via USB, Ethernet, or dual SFP links, in addition to the
VME interface. The Central Card can be configured as a Master that
receives data from up to 8 Slaves, with each Slave managing 8
30-channel front-end cards, allowing a single VME crate to control up
to 1920 channels of the PSEC4 chip.

The organization of the paper is as follows: the integration the PSEC4 ASIC into
a DAQ system is described in Section~\ref{sec:DAQ}. Operating
experience with a 60-channel system at the Argonne APS timing
lab~\cite{RSI_paper} and with a 180-channel system for the
prototype Optical Time Projection 
Chamber (OTPC) in the MCenter Test Beam at Fermilab are presented in
Section ~\ref{sec:Operating}. 
Section~\ref{sec:Conclusion} concludes the paper.

\section{System Overview}
\label{sec:DAQ}

\begin{figure}[!b]
\centering
\includegraphics[width=3.42in]{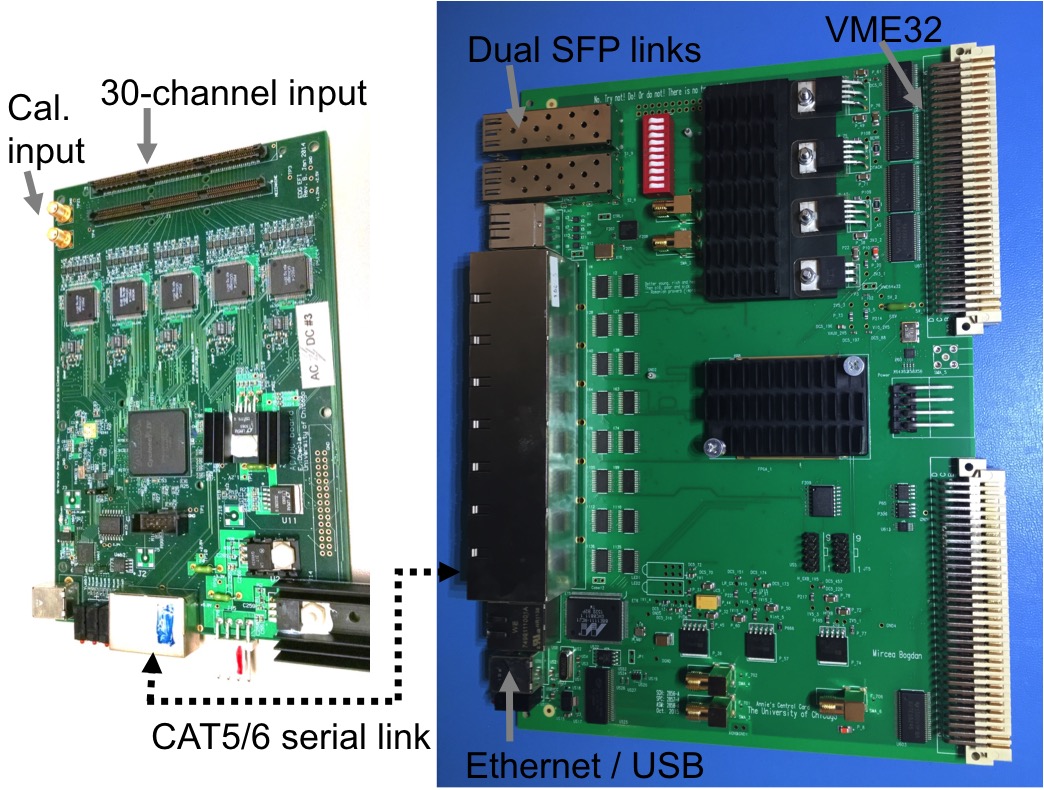}
\caption{System Cards. On the left is the front-end ACquisition and Digitization with pseC4 (ACDC) card.
The ANNIE Central Card (ACC) is shown on the right.
Each back-end ACC manages up to 8 front-end ACDC boards using two network cables per board. Data are transferred using SerDes with a custom protocol at a rate of up to 1.6 Gbps. A single differential pair on the link is dedicated to the system clock, which is sent to a jitter-cleaning chip on ACDC cards.}
\label{fig:Boards}
\end{figure}

The DAQ System, pictured in Figure~\ref{fig:Boards}, 
is based on a back-end digital interface module called the ANNIE\footnote{The Accelerator Neutrino Neutron Interaction Experiment (ANNIE)~\cite{annie}} Central Card (ACC) and a PSEC4-based
digitizing board called the ACquisition and Digitization with pseC4 (ACDC) card. Depending on the firmware installed, ACC can directly service up to eight ACDC boards, or function as a system controller called the Annie Crate Master (ACM). The system clock, as well as trigger and control pulses, are generated in the ACM and passed along the the ACCs in LVDS format via CAT5 cables. In this configuration, one double-width ACM, populated with a 16-port, double-row RJ45 connector can service 8 ACCs. For larger system requirements, this structure can be extended in multiples of eight, in a pyramid scheme.

\begin{figure*}[!t]
\centering
\includegraphics[height=2.2in]{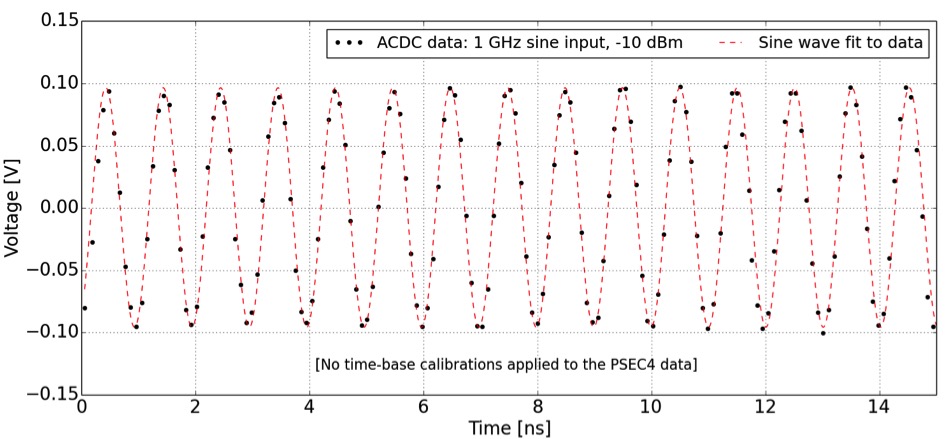}
\includegraphics[height=2.2in]{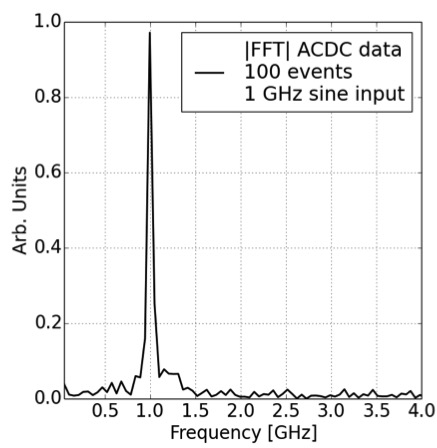}
\caption{1 GHz sine wave captured in one channel of the PSEC4 DAQ system. 
The PSEC4 chip is sampling at 10.24 GSa/sec over the 256 sample-cells per channel. 
The left plot shows the pedestal and linearity-corrected data (black dots) and a sine-wave fit to the data 
(red dotted-line). The right plot is the average FFT magnitude when including 100 recorded sine-waves 
in the transform.}
\label{fig:1GHzSine}
\end{figure*}

For smaller systems, a single-width ACM, made from the same PCB, but populated with an 8-port, single row RJ45 connector can service 4 ACCs, and this type of system can be extended in multiples of 4, again in a pyramid scheme.

The ACCs and the ACM are fitted each with two SPF optical links for data readout. The ACCs can be daisy-chained and data are passed along to the ACM, and from there to a remote PC. Depending on system requirements, the ACM readout can be done via: SFP, CAT5, USB or VME. In the current configuration, data are read via the ACM optical link, and the maximum throughput is 6~Gbps for the entire system.
For higher rates, data can be read in parallel directly from each ACC, for a maximum throughput of 6~Gbps per ACC.

The ACC prototype can function as a typical 6U VME slave, single or double width, or as a stand-alone, 5~V powered unit. The ACC interfaces with each ACDC board via two CAT5 cables. In total, there are 8 LVDS lines to/from each ACDC: four LVDS outputs, including the system clock, and four LVDS inputs. In the ACC printed circuit board the LVDS I/O lines within each ACDC group are length-matched to $\pm$5mm to the Altera 5AGXFB5H4F35C4N FPGA. Each line can run with serialized data rates of up to 1.2~Gbps. The clocks to all ACDC boards bypass the FPGA and are routed to equal length within  $\pm$2.5~mm.

For system integration, the ACC has a separate RJ45 connector supporting three LVDS inputs and one output.
In addition to VME, communication with the ACC can be made via the Front Panel Ethernet and/or USB ports. The module is also provided with six configurable SMA I/Os.

Two ACC prototypes were produced and tested at The University of Chicago. A 1~GHz sine wave recorded with
an ACC prototype and an ACDC card running at 10.24 GSa/sec is shown in Figure~\ref{fig:1GHzSine}.

\section{Operating Experience}
\label{sec:Operating}

\begin{figure}[!ht]
\centering
\includegraphics[trim=1mm 1mm 1mm 1mm, clip=true, angle=0, width=3.4in]{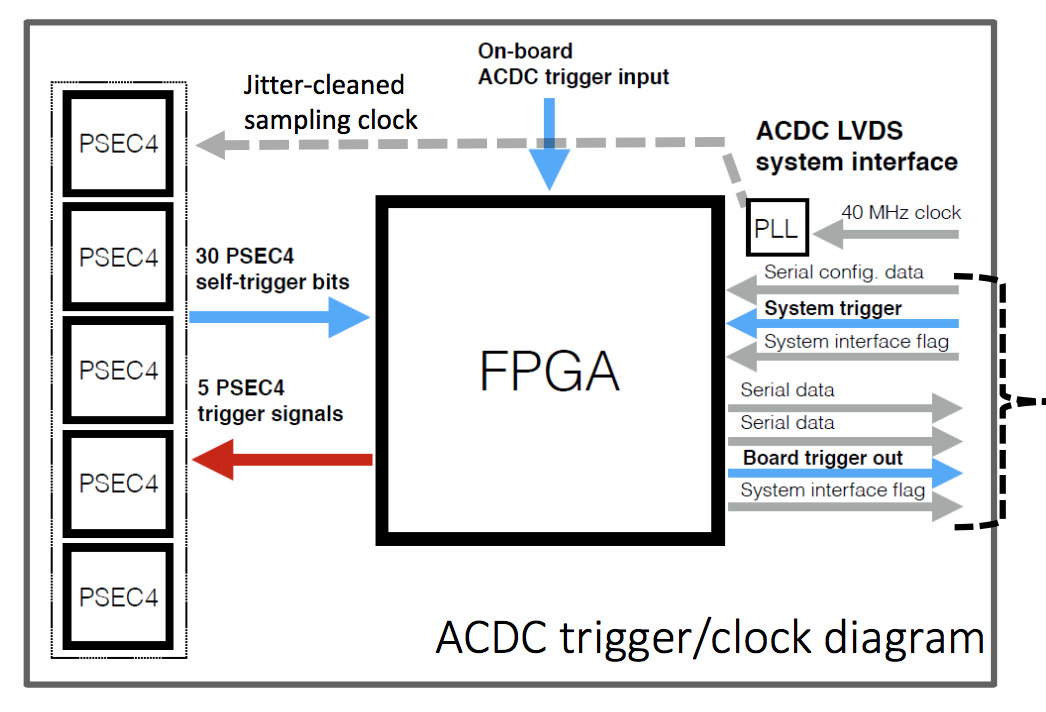}
\caption{ACDC trigger and clock diagram. The 40 MHz clock and 7 additional signals are used to 
communicate with the ACC. An ACDC-local trigger can be formed on the FPGA 
using the output bits of the threshold-level discriminators built into each PSEC4 channel.}
\label{fig:ACDCblock}
\end{figure}

Two prototype systems have been fielded: A 60-channel system used with LAPPD testing and a 180-channel
system used to instrument the OTPC test-beam experiment~\cite{RSI_paper, OTPC_paper}.

The DAQ system can be externally triggered, self-triggered, or built with a coincidence of both.
When running the PSEC4 at a sampling rate of 10.24 GSa/sec, the
waveforms on the PSEC4 capacitor array are over-written in 25 ns intervals. 
To save waveforms when a global trigger has a latency larger than 25 ns, a local 
ACDC trigger can be formed using the built-in threshold discriminators on each channel of the PSEC4 chip. 
In the current design, each PSEC4 has an independent, adjustable threshold level. 
The trigger channel-mask and coincidence window between the local and system trigger are programmable over the serial link from the ACC.

The 180-channel prototype of this data acquisition system was deployed at the 
Fermilab test-beam in a water Cherenkov detector using microchannel plate photo-multiplier tubes (MCP- PMT): 
the ‘Optical Time-Projection Chamber’ (OTPC)~\cite{OTPC_paper}. The system was read out using two Central Cards
programmed in Master-Slave operation. Data were readout over parallel USB 2.0 links to a Linux PC in the test-beam enclosure.

\begin{figure}[!ht]
\centering
\includegraphics[trim=1mm 10mm 60mm 10mm, clip=true, angle=0, width=3.2in]{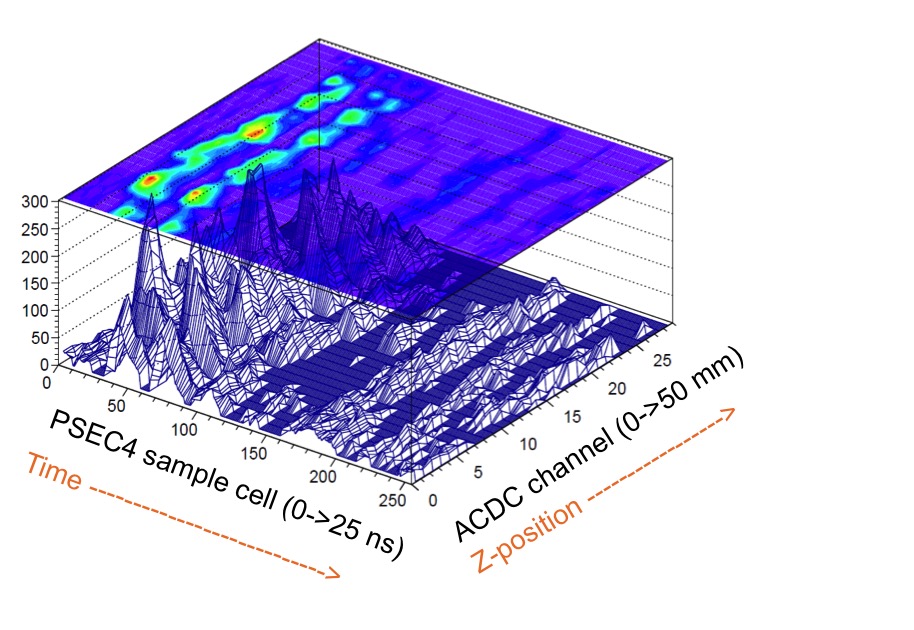}
\caption{OTPC event at a single MCP-PMT. The signal from a number of Cherenkov photons 
from a relativistic muon detected at a 
microstrip-readout MCP-PMT, self-triggered and recorded at a 30-channel ACDC board. Each PSEC4 channel
reads out a 1.7~mm wide microstrip, with 30~microstrips covering the MCP-PMT area. When a photo-electron 
signal is received at the anode, the two pulses recorded in a channel are the direct and open-end reflected pulse on the anode. The time difference between the two pulses gives the reconstructed position of the photon
along the microstrip.
The vertical, z-axis, is in units of PSEC4 ADC counts (1 ADC count$\sim$0.3~mV). 
More details on the OTPC can be found in Ref.~\cite{OTPC_paper}}
\label{fig:OTPC}
\end{figure}

In the OTPC system, the ACDC boards formed level-0 triggers using the trigger bits from PSEC4-discriminated single photon MCP-PMT pulses. A 1.8 GHz 20~dB pre-amp board was designed to increase the single-photon trigger efficiency. Single photo-electron pulses were typically between 50 and 80~mV in amplitude and $\sim$1 ns FWHM after the pre-amp board.
Data were read out from the ACDC cards when a global trigger was registered at the Central Card.
An OTPC event (at a single MCP-PMT) from a relativistic muon traversing the water detector is shown in 
Figure~\ref{fig:OTPC}.

\section{Conclusion}
\label{sec:Conclusion}
We have built a modular DAQ system around the 5-15 GS/sec PSEC4 waveform
sampling ASIC. The system consists of two levels, a front-end card
that supports 5 of the 6-channel PSEC4 chips, and a back-end `Central'
VME board that controls 8 of the front-end cards, for a total of 240
channels. The Central card can be configured as a master controlling 8
other Central cards as Slaves, for a total of 1920 channels. The
Central card supports  USB, Ethernet, or dual 6-Gb/sec SFP links,
in addition to the VME interface.

\section{Acknowledgments} We thank Mary Heintz, 
Jean-Francois Genat, Herv{\'{e} Grabas, Kurtis Nishimura, Harold
  Sanders, Fukun Tang, and Gary Varner for their essential contributions.
 \label{sec:Acknowledgements}

\end{document}